\documentclass[reprint,aps, superscriptaddress,nofootinbib]{revtex4-1}

\usepackage{graphicx}
\usepackage{dcolumn}
\usepackage{bm}

\usepackage{amsmath}
\usepackage{amssymb}
\usepackage{amsfonts}
\usepackage{amsthm}
\usepackage{color}
\usepackage{subfig}
\usepackage{MnSymbol}

\begin{document}

\captionsetup{justification=raggedright,singlelinecheck=false}

\title{Numerical simulations  of quantum clock for measuring tunneling times}

\date{\today}

\author{Fumika Suzuki}

\affiliation{%
Theoretical Division, Los Alamos National Laboratory, Los Alamos, New Mexico 87545, USA
}

 \affiliation{%
Center for Nonlinear Studies, Los Alamos National Laboratory, Los Alamos, New Mexico 87545, USA
}

  \affiliation{%
ISTA (Institute of Science and Technology Austria), Am Campus 1, 3400 Klosterneuburg, Austria
}

\email{fsuzuki@lanl.gov}

\author{William G. Unruh}

  \affiliation{%
Department of Physics and Astronomy, University of British Columbia, Vancouver, British Columbia V6T 1Z1, Canada
}

\affiliation{Institute for Quantum Science and Engineering, Texas A\&M University, College Station, Texas 77843, USA}
  
 \email{unruh@physics.ubc.ca}
  \begin{abstract}
  
We numerically study two methods of measuring tunneling times using a quantum clock. In the conventional method using the Larmor clock, we show that the Larmor tunneling time can be shorter for higher tunneling barriers. In the second method, we study the probability of a spin-flip of a particle when it is transmitted through a potential barrier including a spatially rotating  field interacting with its spin. According to the adiabatic theorem, the probability depends on the velocity of the particle inside the barrier. It is numerically observed that the probability increases for higher barriers, which is consistent with the result obtained by the Larmor clock. By comparing outcomes for different initial spin states, we suggest that one of the main causes of the apparent decrease in the tunneling time can be the filtering effect occurring at the end of the barrier.
\end{abstract}

\maketitle


\section{Introduction}

Measurement of time is often ambiguous in quantum mechanics due to the absence of time operator \cite{time3, time2, time, aoki}. In particular, the problem of quantum tunneling time, i.e., ``How long does quantum tunneling take?", has been a long-standing controversial issue of quantum mechanics \cite{bohm, wigner, review, review2,davis, review3, review4, mac, hart, krenz}. Quantum tunneling has been studied in various fields, including superconductors \cite{superconducting}, spintronics \cite{TMR}, micromaser fields \cite{scully}, nuclear fusion \cite{nuclear}, biological or chemical processes \cite{arndt, chem2}, composite particle  dynamics \cite{comp, comp2,comp3, comp4, comp5} and relativistic quantum mechanics \cite{rel,rel2,rel3}. Many attempts have been made to define tunneling times, including the phase times \cite{bohm, wigner}, the dwell time \cite{smith}, the Larmor time \cite{larmor}, or by using the time-dependent potential barrier \cite{landauer}, paths integrals \cite{sok} and weak measurements \cite{weak, weak2}. It has now become possible to address the problem experimentally using strong field tunneling ionization \cite{exp2} or ultracold atoms \cite{exp, exp3}. One of the simplest possible methods to measure tunneling times can be to compare the difference in the time of arrival of a wave packet with and without a tunneling barrier \cite{mac,hart}. Precise measurement by this method requires that both the initial wave packet and the wave packet after tunneling through a barrier have small uncertainty in position. However, this results in large uncertainty in momentum, making it difficult to restrict the modes of the wave packet to have energies less than the barrier height. It is generally known that the position uncertainty of the wave packet should be greater than the width of the barrier in order to simultaneously satisfy the conditions that most modes in the wave packet have  energies less than the barrier height and that the transmission probability of the particle through the barrier is reasonably large. As a result, the uncertainty of the measurement can be greater than the measured tunneling time by this method. Another method that has been recently implemented by an experiment is to use the Larmor clock \cite{larmor, exp, exp3}. In this method, a quantum system such as spin is attached to a particle as a clock. The clock then could be used to measure tunneling times when it is made to run only within the barrier \cite{time3, larmor}. It was observed that the Larmor time, interpreted as the tunneling time, appears to become shorter as the barrier height increases \cite{exp, exp3}. Although many analytical studies have been done on tunneling times, there are few numerical calculations on the subject. In this paper, we numerically study the use of quantum clocks for measuring tunneling times. In particular, we focus on the impact of the measurement-induced backaction and the filtering effect of the barrier, i.e., the preferential transmission of high momentum modes by the barrier. In addition to the method using the Larmor clock, we introduce a new method using the adiabatic theorem to investigate time-of-flight and tunneling times. Comparing the two methods may clarify the differences in behaviors caused by the backaction and the filtering effect.

This paper is organized as follows: In Sec. II, we review the measurements of time-of-flight and tunneling times using the Larmor clock. We show that the decrease of the Larmor time for higher barriers is observed in the numerical simulation of the wave packet dynamics. The numerical results obtained are similar to those observed experimentally in \cite{exp, exp3}. In Sec III, we introduce a new method of investigating time-of-flight and tunneling times using the adiabatic theorem. The adiabatic theorem    \cite{adiabatic2} has been applied to a wide range of contexts in quantum mechanics, such as quantum phase transitions \cite{kibblezurek, kibblezurek2, kibblezurek3, kibblezurek4, kibblezurek5, kibblezurek6, kibblezurek7}, geometric phase \cite{berry}, quantum computations \cite{quantumcomp}, chemical reactions \cite{chem} and atomic or molecular collision theory \cite{collision, collision2}. We consider the adiabatic theorem for a spin of a particle that propagates through the region with a gradual rotation of the direction of the field interacting with the spin. The model could be relevant to electronic transport through a  domain wall in a ferromagnet \cite{domain, domain2, domain3} and spin transistor action \cite{trans, trans2}. The high probability of a spin flip of the particle after transmission can indicate that the particle traversed the barrier non-adiabatically. It is observed that the particle transmitted through the higher barrier exhibits a higher probability of a spin flip, which is consistent with the results discussed in Sec. II. By performing numerical simulations with different initial spin states, we explore the relevance of the backaction and the filtering effect to these observations. Finally, we conclude that the filtering effect occurring at the end of the barrier can be one of the main causes of the apparent decrease in the tunneling time with increasing potential height (Sec. IV).

\section{Larmor clock for tunnelling times}

In this section, we review the measurements of time-of-flight and tunneling times using the Larmor clock by performing numerical simulations. 
The Larmor clock can be described by the dynamics of a spin-$1/2$ particle whose spin experiences Larmor precession in the region where it interacts with the  magnetic field, i.e., $|y|\leq D$. The Hamiltonian is given by
\begin{eqnarray}\label{ham00}
H=H_0+H_{SF}=\frac{\hat{k}^2 }{2m}-\frac{\omega_0}{2} g(y)\sigma_z
\end{eqnarray}
where $m$ is the mass of the particle, $\omega_0$ is the coupling constant, $\sigma_z$ is Pauli matrix, and $g(y)=1$ for $|y|\leq D$ and $g(y)=0$ otherwise.

We assume that the particle travels in the $\hat{\mathbf{y}}$-direction and its spin is initially polarized in  the $\hat{\mathbf{x}}$-direction. When the particle initially starts at $y\ll -D$, the time-of-flight in the region where $|y|\leq D$ is measured using the Larmor precession. On the measurement of the spin state of the particle at $y\gg D$, we define
\begin{eqnarray}
\tau_y =\frac{1}{\omega_0}\mbox{arctan}-\frac{\langle S_y\rangle}{\langle S_x\rangle}, \quad \tau_z =\frac{1}{\omega_0}\mbox{arctan}\frac{\langle S_z\rangle}{\sqrt{\langle S_x\rangle^2+\langle S_y\rangle^2}}\nonumber\\
\end{eqnarray}
where $\langle S_x\rangle$, $\langle S_y\rangle$ and $\langle S_z\rangle$ are the expectation values of the spin component. $\tau_y$ represents the time-of-flight of the particle in the region where $|y|\leq D$ and $\tau_z$ is associated with the measurement-induced backaction caused by the interaction of the magnetic field with the spin of the particle.

As an example, we prepare the initial wave packet of the particle, $|\Psi_{\rm in}\rangle =\psi_0 (y)|s\rangle$ where $|s\rangle =|\uparrow\rangle$ is the spin-up state in the $\hat{\mathbf{x}}$-direction, and
\begin{eqnarray}\label{wavepacket}
\psi_0 (y)=\frac{1}{(2\pi\sigma_y^2)^{1/4}}\exp \left(-\frac{(y-y_0)^2}{4\sigma_y^2}+ik_0 y\right)
\end{eqnarray}
with the uncertainty of position $\sigma_y$ and $k_0$ representing the momentum in the $\hat{\mathbf{y}}$-direction.

\begin{figure}
{%
\includegraphics[clip,width=\columnwidth]{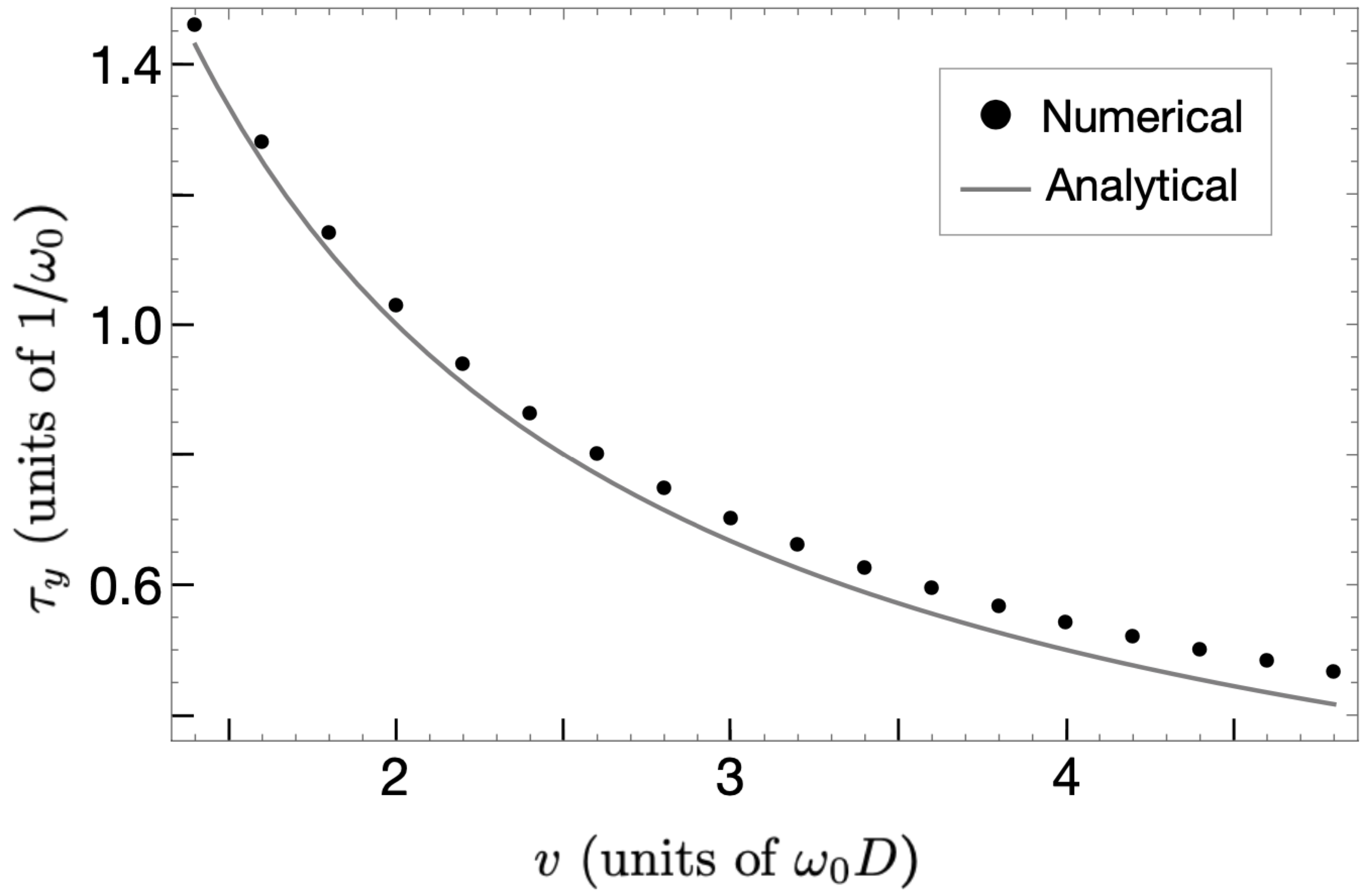} 
}
\caption{$\tau_y$ with different initial velocities obtained numerically (black circles). The solid black line represents the analytical estimate $\tau_y = 2D/v_0$ for comparison.}
\label{fig1}
\end{figure}

\begin{figure}
{%
\includegraphics[clip,width=0.8\columnwidth]{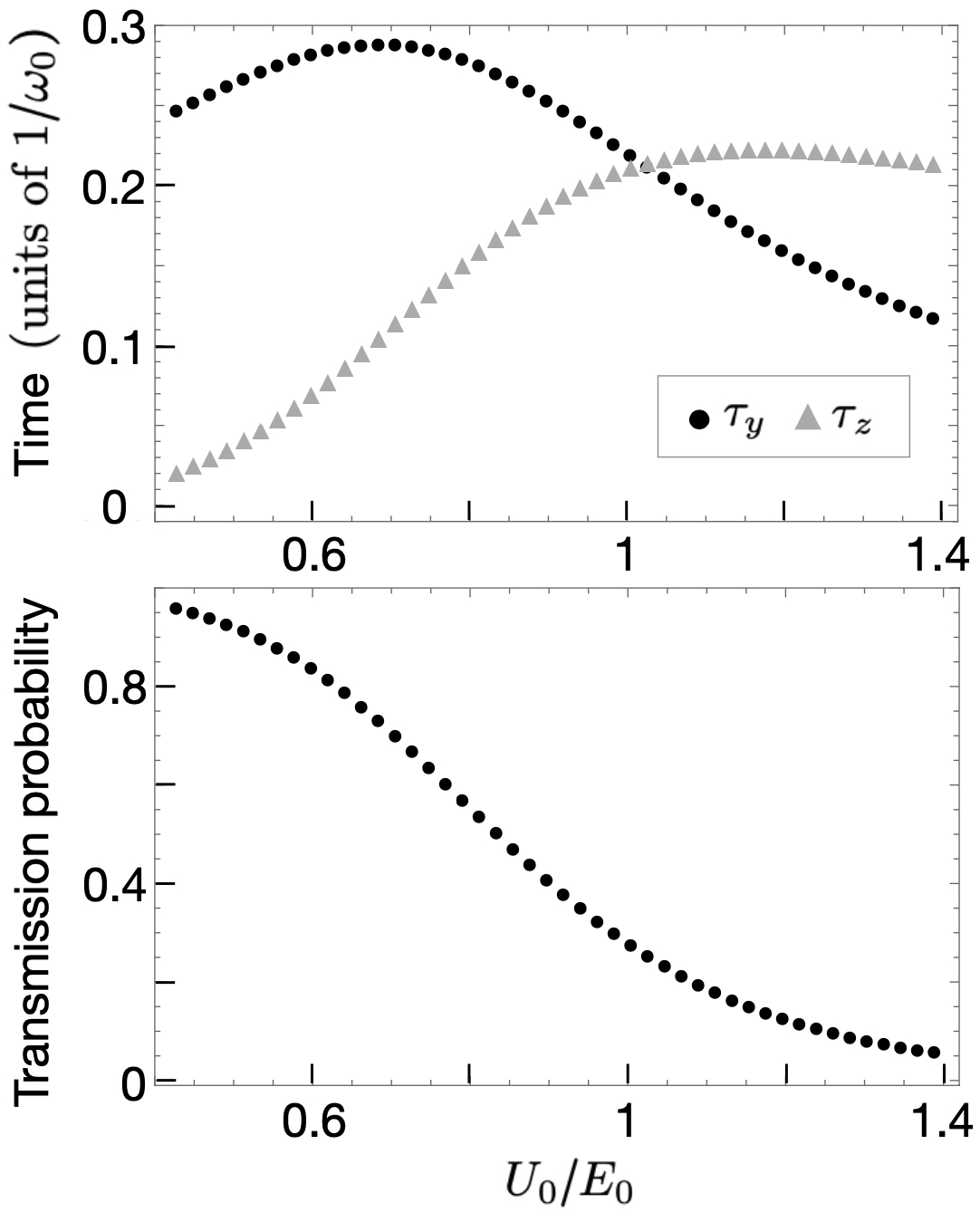} 
}
\caption{$\tau_y$, $\tau_z$ (upper panel), and transmission probability  (lower panel) as the function of $U_0/E_0$ obtained numerically.}
\label{fig2}
\end{figure}

The particle is initially located at $y_0=-9.5D$ at time $t=0$ and we choose $\sigma_y/D=1$.  We numerically solve the time-dependent Schr\"{o}dinger equation with $H$ (\ref{ham00}) using finite-difference methods to obtain the time evolution of the wave packet. After the propagation of the particle through the region $|y|\leq D$, we obtain the wave packet  $|\Psi_{\rm out}\rangle =e^{-iHt}|\Psi_{\rm in}\rangle$. We evaluate  $\tau_y$ from the wave packet arriving at $y>D$ at $t= 15D/v_0$ with different $v_0$ such that $\omega_0/E_0\in [0.01, 0.13]$ where $v_0=k_0/m$ and $E_0=k_0^2/2m$. Fig. \ref{fig1} shows $\tau_y$ obtained by the above numerical method (black circles). Since the expectation value of the particle velocity is given by $\langle k\rangle /m=v_0$, the time-of-flight of the particle over the region $|y|\leq D$ is analytically estimated as $2D/v_0$ (solid black line). It can be seen from the figure that the Larmor clock makes it possible to measure time-of-flight. However, this measurement has limitations. In order to have a good resolution of time, a large energy transfer between the clock and the translational motion of the particle is necessary. This energy transfer modifies the dynamics of the particle. In \cite{time3}, it was estimated that $\omega_0 \sim 1/\Delta T$ where $\Delta T$ is the time resolution of the clock. Assuming $\omega_0 \ll E_0$ so that the effect of measurement is small, the lower limit of the time resolution of the clock is given by $\Delta T \gg 1/E_0$. It has been further argued that this lower limit imposes a limitation on the accuracy of the measurement of the particle velocity over a distance $2D$. Since  $v \sim 2D/T$ where $T$ is the time-of-flight, $\Delta v \sim v^2 \Delta T /2D \gg 1/2Dm$ or $\Delta k \gg 1/2D$. Therefore only measurements on the particle with $k_0 \gg 1/2D$ can have reasonable accuracy. These are inherent limitations of time-of-flight measurements through quantum clock. Nevertheless, the Larmor clock is commonly used to investigate tunneling times. In the following, we discuss the results which can be obtained when the tunneling time is measured using the Larmor clock despite these limitations.

In order to study the quantum tunneling problem, we introduce the potential barrier:
\begin{eqnarray}
H=\frac{\hat{k}^2 }{2m}-\frac{\omega_0}{2} g(y)\sigma_z + U(y)
\end{eqnarray}
where $U(y)$ is the rectangular potential barrier such that $U(y)=U_0$ for $|y| \leq D$ and $U(y)=0$ otherwise.

We prepare the wave packet $|\Psi_{\rm in}\rangle$ as above with $y_0=-50D$ and $|s\rangle=|\uparrow\rangle$. We choose $\sigma_y/D=10$ and $\omega_0/E_0=0.1$. As in the previous case, the wave packet propagates through the region $|y|\leq D$. $\tau_y$ and $\tau_z$ are measured from the wave packet arriving at $y>D$ at $t=120D/v_0$. Fig. \ref{fig2} shows $\tau_y$, $\tau_z$ (upper panel), and transmission probability of the particle through the barrier (lower panel) as the function of $U_0/E_0$ obtained numerically. Since the velocity inside the barrier can be approximated by $v'\sim \sqrt{2m(E_0-U_0)}/m$ when $U_0/E_0\ll 1$, it can be seen that $\tau_z \sim 2D/v'$ increases with $U_0/E_0$ in the regime. However, as $U_0$ approaches $E_0$ and exceeds it, $\tau_y$ starts to decrease. On the other hand, $\tau_z$ generally tends to increase with  $U_0/E_0$. It was confirmed that the expectation value of the kinetic energy $\langle E\rangle$ for the wave packet arriving at $y>D$ gives $\langle E \rangle < U_0$ when $U_0/E_0 >1$. Therefore, most modes of the transmitted wave packet have experienced the tunneling effect in the regime. Similar results were obtained in the experiment attempting to measure the tunneling time using the Larmor clock \cite{exp, exp3}. The decrease of $\tau_y$ in the tunneling regime was interpreted as the tunneling taking less time for higher barriers. In the next section, we introduce an alternative method to investigate  time-of-flight and tunneling times using the adiabatic theorem and explore the possible causes of these results.

\section{Adiabatic theorem for tunnelling times}

In the previous section, the backaction $\tau_z$ results in different transmission probabilities for spin-up and spin-down states in the $\hat{\mathbf{z}}$-direction, while the initial spin state is prepared in the spin-up state in the $\hat{\mathbf{x}}$-direction. In this section, we introduce a new method to investigate time-of-flight and tunneling times using the adiabatic theorem where the initial spin state can be set to the spin-up or spin-down state in the $\hat{\mathbf{x}}$-direction, and the spin-field interaction causes different transmission probabilities for these spin states in the $\hat{\mathbf{x}}$-direction as well. For this reason, the following method may clarify the relationship between outcomes of tunneling time measurements and measurement-induced backaction.

We consider the Hamiltonian for the spin-$1/2$ particle whose spin is interacting with  the spatially rotating  field\footnote{The situation relevant to our toy model could be a conduction electron locally exchange coupled to electrons in a fixed configuration responsible for the magnetization of domains (separated by a domain wall). }:
\begin{eqnarray}\label{ham0}
H=H_0+ H_{SF}=\frac{\hat{k}^2 }{2m}+\frac{\omega_0}{2} \vec{f} \cdot \vec{\sigma} 
\end{eqnarray}
where $\vec{f}$ represents the direction of the field, and $\vec{\sigma}=(\sigma_x,\sigma_y,\sigma_z)$ are the Pauli matrices. We assume that the particle travels in the $\hat{\mathbf{y}}$-direction.

We choose
\begin{eqnarray}\label{mag}
\vec{f}=\begin{cases}
      -\sin \frac{\pi y}{2D} \hat{\mathbf{x}}+  \cos \frac{\pi y}{2D}\hat{\mathbf{y}}, & \text{for}\ |y|<D\\
      - \mbox{sgn}(y)\hat{\mathbf{x}}, & \text{for}\ D\leq |y|\leq L\\
       \vec{0}, & \text{for}\ |y|> L
    \end{cases}
\end{eqnarray}
where $L\geq D$.

In this choice, the field rotates in the $x-y$ plane. However, the following arguments are equally applicable to other choices (e.g., the rotation of the field in the $x-z$ plane). We have
\begin{eqnarray}
H_{SF}=\begin{cases}\frac{\omega_0}{2}\begin{pmatrix}0&-i\exp (-i\frac{\pi y}{2D})\\i\exp \left(i\frac{\pi y}{2D}\right)&0\end{pmatrix}, & \text{for}\ |y|<D\\
-\frac{\omega_0}{2}\begin{pmatrix}0&\mbox{sgn}(y)\\\mbox{sgn}(y)&0\end{pmatrix}, & \text{for}\ D\leq |y|\leq L\\
0_{2\times 2}, & \text{for}\ |y|> L
\end{cases}.\nonumber\\
\end{eqnarray}

\begin{figure}
{%
\includegraphics[clip,width=\columnwidth]{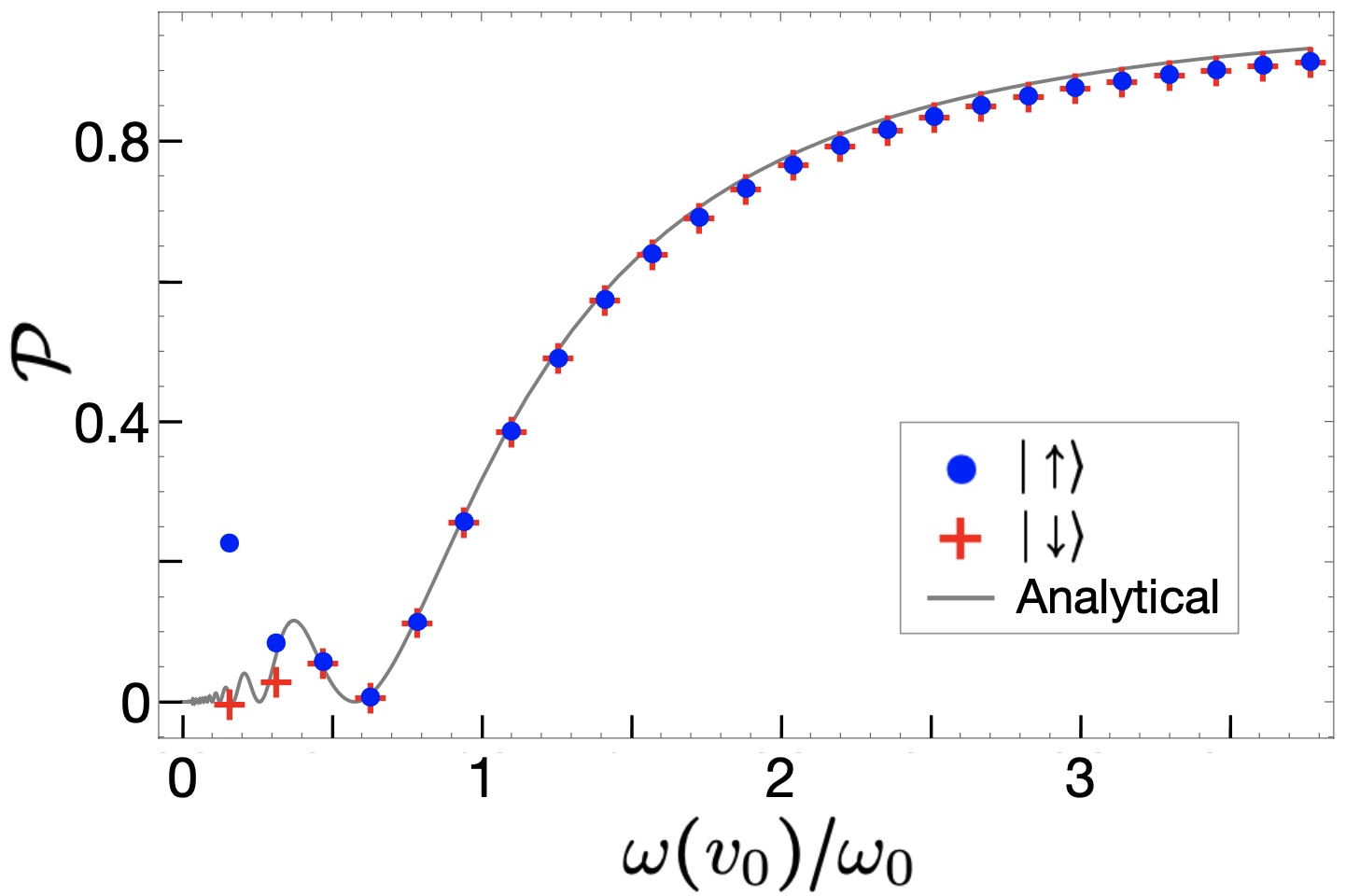} 
}
\caption{The probability of a spin flip $\mathcal{P}$ obtained numerically. The initial spin state is prepared in $|\uparrow\rangle$ (blue circles) and $|\downarrow\rangle$ (red crosses) respectively.  The solid grey line is given by the analytical expression Eq. (\ref{spinflip}) for comparison.}
\label{fig3}
\end{figure}

\begin{figure}
{%
\includegraphics[clip,width=\columnwidth]{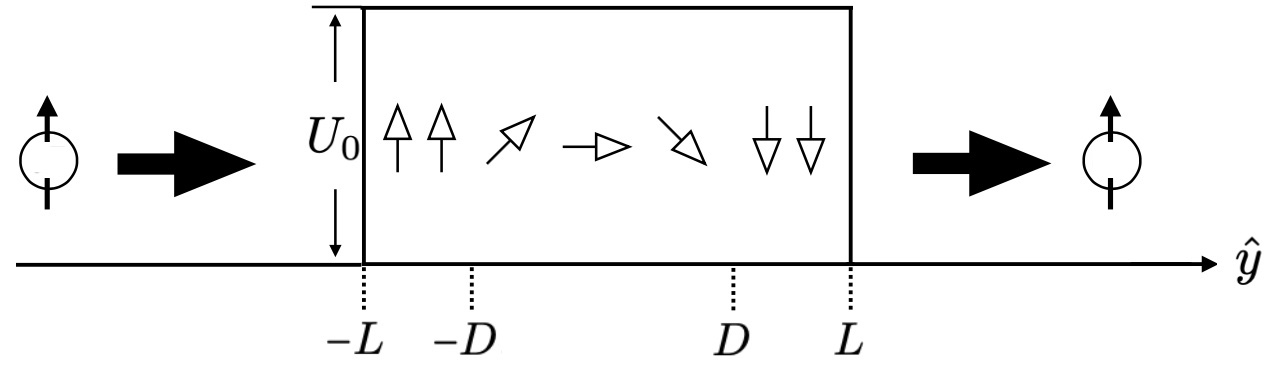} 
}
\caption{A model used for investigating quantum tunneling dynamics using the adiabatic theorem. The particle propagates through the potential barrier including the spatially rotating  field.}
\label{fig4}
\end{figure}

\begin{figure*}
{%
\includegraphics[clip,width=1.7\columnwidth]{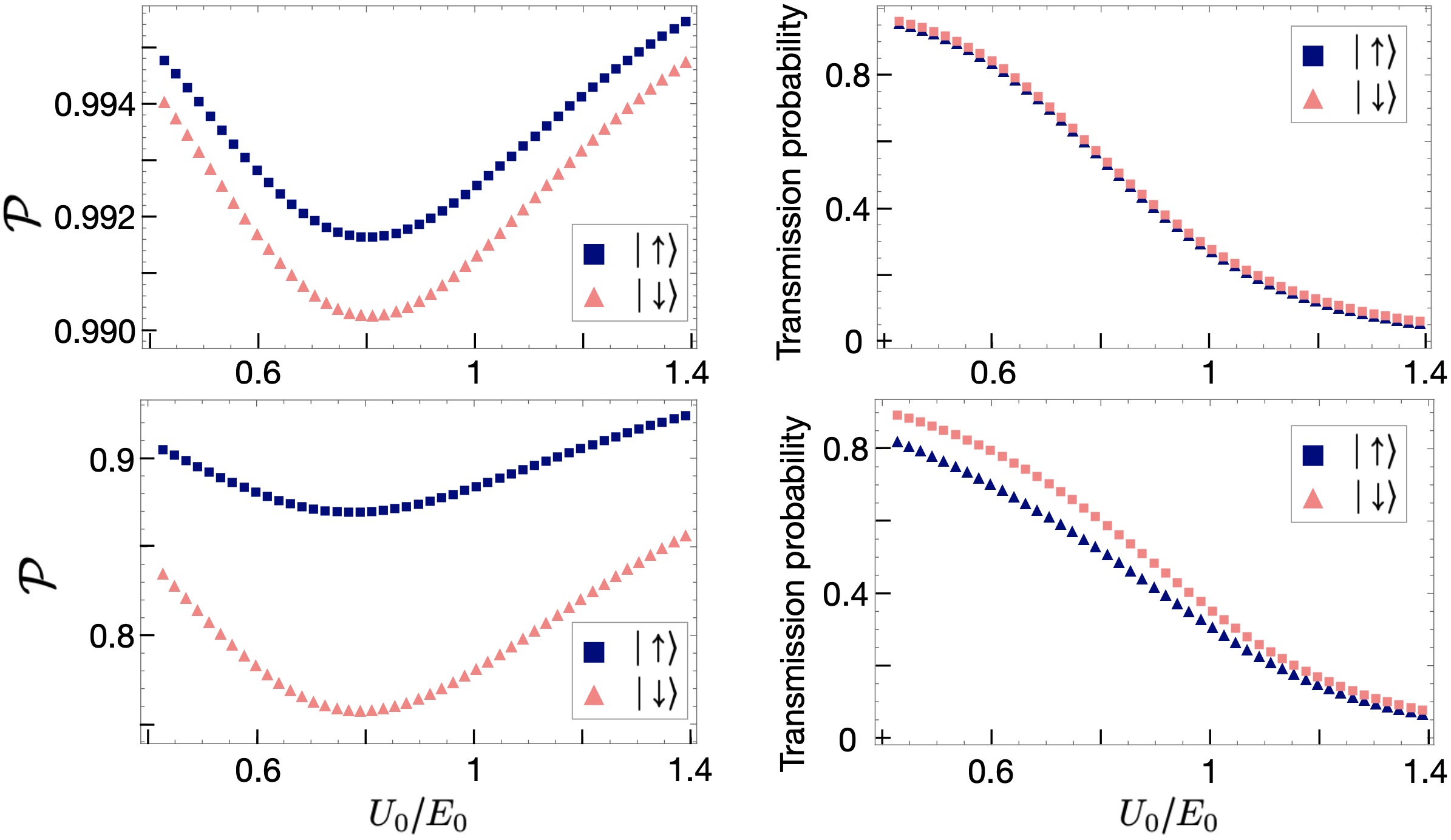} 
}
\caption{The probability of a spin flip $\mathcal{P}$ (left) and transmission probability (right) as the function of $U_0/E_0$ obtained numerically. The initial spin state is prepared in $|\uparrow\rangle$ (navy squares) and $|\downarrow\rangle$ (pink triangles) respectively. $\omega_0/E_0=0.1 $ (upper panel) and $\omega_0/E_0 =0.5$ (lower panel).}
\label{fig5}
\end{figure*}

It is assumed that $\vec{f}=\vec{0}$ when $|y|> L$. However, the discussions below also apply to the case where the uniform  field $\vec{f}= -  \mbox{sgn}(y)\hat{\mathbf{x}}$ exists for $|y| > L$. The particle initially starts at $y\ll -L$ with the spin-up or spin-down state along the field at $y=-D$. Its initial wave packet is given by (\ref{wavepacket}) as before. We calculate the probability of a spin flip after the propagation through the region $|y|\leq D$.  If the particle propagates such that $\langle y(t)\rangle = v_0 t$ where $v_0=k_0/m$, $H_{SF}(y(t))$ appears as the time-dependent Hamiltonian with the field rotating at the angular velocity $\omega(v_0)=\pi v_0/2D$ from the point of view of the spin degrees of freedom. Therefore the approximate time-dependent Schr\"{o}dinger equation for a spin state $\chi (t)$, $i\hbar\frac{\partial}{\partial t}\chi (t)\approx H_{SF} (t)\chi (t)$ gives the familiar problem of a particle with a spin in the rotating  field, and the probability of a spin flip can be calculated as \cite{book}
\begin{eqnarray}\label{spinflip}
\mathcal{P}=\left[\frac{1}{\sqrt{1+(\omega_0 /\omega (v_0) )^2}}\sin \left(\frac{\pi}{2}\sqrt{1+(\omega_0/\omega (v_0))^2}\right)\right]^2.\nonumber\\
\end{eqnarray}

This indicates that the adiabaticity is maintained when the velocity is small and the rotation of the  field is slow in the perspective of the spin state so that  the spin can track the reorientation of the  field. In other words, $\mathcal{P}\rightarrow 0$ when  $\tau_D \gg \tau_0$ where $\tau_D=1/\omega (v_0)$ is the characteristic time for a change in $H_{SF}$  and $\tau_0=1/\omega_0$.

In the following, we choose $\sigma_y/D=1$ and $L/D=2$. The particle is initially located at  $y_0=-9.5D$ at time $t=0$. We numerically solve the time-dependent Schr\"{o}dinger equation with $H$ (\ref{ham0}). After the propagation of the particle through the region $|y|\leq D$, we obtain the wave packet  $|\Psi_{\rm out}\rangle =e^{-iHt}|\Psi_{\rm in}\rangle =\sum_{s=\uparrow,\downarrow}\psi_{s}(y)|s\rangle$. We evaluate the probability of a spin flip at $t= 15D/v_0$ by normalizing the wave packet at $y\geq D$, i.e., $\mathcal{P}=\int_{D}^{\infty}\psi_{s}^{*}(y)\psi_{s}(y)dy / \sum_{s'=\uparrow,\downarrow}\int_{D}^{\infty}\psi_{s'}(y)^{*}\psi_{s'}(y)dy$ where $s=\downarrow$ or $\uparrow$ when the initial spin state starts with $|\uparrow\rangle$ and $|\downarrow\rangle$ respectively. This is the probability for finding the spin state $|\downarrow\rangle$ or $|\uparrow\rangle$ along the direction of the field at $y= D$ when the state is initially prepared in $|\uparrow\rangle$ or $|\downarrow\rangle$ respectively along the direction of the field at $y= -D$ before the propagation.

By repeating the above numerical computations with different $v_0$ so that  $\omega_0/E_0 \in [0.01, 6.2]$ where $E_0=k_0^2/2m$, we obtain the result in Fig. \ref{fig3}. When $\omega(v_0)/\omega_0 \gtrsim 1$, it can be seen that the result agrees well with the analytical plot from Eq. (\ref{spinflip}) represented by the solid grey line. This confirms that, in the regime of the weak field, the probability of a spin flip for the particle propagating through the spatially rotating field can be estimated by the adiabatic theorem where the nonadiabaticity is determined by the velocity of the particle $v_0$. On the other hand, the strong field modifies the dynamics of the particle significantly. As a result, the time evolution of the particle becomes different depending on the initial spin state, and the numerical result deviates from the analytical estimate  (\ref{spinflip}) when $\omega (v_0)/\omega_0 <1$.

In this model, the spin interacting with the field can measure time-of-flight $T\sim \pi \tau_D$ by observing the velocity of the particle. However, we have $\mathcal{P}\rightarrow 0$ and $\mathcal{P}\rightarrow 1$ when $\tau_D \gg \tau_0$ and $\tau_D \ll \tau_0$ respectively. Therefore, for a reasonable resolution of time, it is necessary to have  $T \sim \tau_0$. This indicates that the energy transfer between the spin and the translational motion of the particle should be large when $T$ is small. Since the uncertainty of the momentum of the particle becomes $\Delta k \sim  \omega_0/v_0 \sim 1/2D$ by the energy transfer, only measurements on the particle with $k_0 \gg 1/2D$ would have reasonable accuracy. Therefore, this method also cannot avoid the inherent limitations, similar to the previous method using the Larmor clock. However, in this method, both time-of-flight measurement and backaction occur to the spin-up and spin-down states in the $\hat{\mathbf{x}}$-direction, which may clarify the relation between the two more directly. By comparing outcomes for different initial spin states, we investigate the effect of the backaction when the method is applied to the study of tunneling times.\\

Let us consider the situation where there exists  the potential barrier in addition to the  field (Fig. \ref{fig4}).  The Hamiltonian can be written as
\begin{eqnarray}\label{ham2}
H=\frac{\hat{k}^2}{2m}+\frac{\omega_0}{2} \vec{f} \cdot \vec{\sigma}  +U(y)
\end{eqnarray}
where we introduce the rectangular potential barrier such that $U(y)=U_0$ for $|y| \leq L$ and $U(y)=0$ otherwise.

When $E_0 > U_0 \pm \omega_0/2$, it is known that the velocity inside the barrier can be approximated by $v' \sim  k_{\uparrow, \downarrow}/m=\sqrt{2m(E_0-U_0\mp \omega_0/2)}/m$
where $k_{\uparrow}$ and $k_{\downarrow}$ correspond to the momentum of the particle with spin-up state and spin-down state respectively. However, $k_{\uparrow, \downarrow}$ inside the barrier in the tunneling regime $E_0 < U_0 \pm \omega_0/2$ is imaginary. We use this model to study the nonadiabaticity of the propagation of the particle in the tunneling regime. Fig. \ref{fig5} shows the probability of a spin flip $\mathcal{P}$ of the transmitted wave packet after the propagation through the potential barrier (left) and transmission probability of the particle through the barrier (right) obtained numerically using the Hamiltonian (\ref{ham2}) and the initial wave packet $|\Psi_{\rm in}\rangle=\psi_0 (y)|s\rangle$  (\ref{wavepacket}) with $|s\rangle =|\uparrow\rangle$ (navy squares) or with $|s\rangle =|\downarrow\rangle$ (pink triangles). Here $\mathcal{P}=\int_{L}^{\infty}\psi_{s}^{*}(y)\psi_{s}(y)dy / \sum_{s'=\uparrow,\downarrow}\int_{L}^{\infty}\psi_{s'}(y)^{*}\psi_{s'}(y)dy$ with $|\Psi_{\rm out}\rangle =e^{-iHt}|\Psi_{\rm in}\rangle =\sum_{s=\uparrow,\downarrow}\psi_{s}(y)|s\rangle$ and $\mathcal{P}$ is plotted as the function of $U_0/E_0$. We chose $\sigma_y/D=10$, $L/D=1$, and $\omega_0/E_0=0.1$ in the upper panels and  $\omega_0/E_0=0.5$ in the lower panels respectively. It was confirmed that the expectation value of the kinetic energy $\langle E\rangle$ for the wave packet arriving at $y>L$ gives $\langle E \rangle <U_0\pm\omega_0/2$ (for $|s\rangle =|\uparrow\rangle$ and $|\downarrow\rangle$ respectively) when $U_0/E_0>1$ with $\omega_0/E_0=0.1$, while the condition is satisfied when $U_0/E>1$ for $|s\rangle =|\uparrow\rangle$ and $U_0/E_0>1.27$ for $|s\rangle =|\downarrow\rangle$ with $\omega_0/E_0=0.5$.   This indicates that most modes of the transmitted wave packet have undergone a tunneling process in these regimes. When $L$ and $D$ are large and the interaction time between the field and the spin can be long, it is possible to measure the nonadiabaticity of the propagation with a small $\omega_0$. However, the tunneling probability becomes extremely low in the situation. In order to obtain a reasonably high tunneling probability, the length of the barrier $2L$ should be around $1/\kappa$ where $\kappa=\sqrt{2m(U_0-E_0)}$. With this length, an appropriate resolution for the measurement can be obtained with $\omega_0$ which causes the momentum transfer  $\Delta k\sim 1/2D \geq  \kappa$. In other words, the energy transfer $\Delta E\geq  \kappa^2/2m$ is necessary. Comparing the upper and lower left panel for the probability of a spin flip, it can be seen that $\mathcal{P}$ is more dependent on $U_0/E_0$ in the lower panel. This makes it appear that the case of the lower panel allows a more precise determination of the nonadiabaticity of the propagation. However, a large $\omega_0$  has a great effect on the dynamics of the particle instead and can alter it significantly. This can be confirmed by the larger difference in the transmission probability depending on the initial spin state in the lower right panel. Since the transmission probability of the spin-down state is higher than that of the spin-up state, $\mathcal{P}$ is higher when the initial spin state is prepared in the spin-up state $|s\rangle=|\uparrow\rangle$, as can be seen in the left panels.  The difference in $\mathcal{P}$ at each $U_0/E_0$ between the initial state $|s\rangle=|\uparrow\rangle$ and $|\downarrow\rangle$ is approximately the same as the difference in the transmission probability between these states at each $U_0/E_0$. This suggests that the backaction is mainly responsible for this difference by causing spin-dependent transmissions. In both the upper and lower panels, $\mathcal{P}$ decreases as $U_0/E_0$ increases when $U_0/E_0$ is sufficiently smaller than $1$ since the velocity of the particle decreases inside the potential barrier. However, as $U_0$ approaches $E_0$ and exceeds it to enter the tunneling regime ($U_0/E_0 \gtrsim1$), $\mathcal{P}$ starts to increase again.  Remarkably, this behavior can be seen in both initial spin states  $|\uparrow\rangle$ and $|\downarrow\rangle$. This indicates that the behavior can be attributed to the filtering effect rather than the spin dependence of the transmission probability due to the backaction. It may be understood as follows. The study of tunneling dynamics generally requires the initial preparation of a spatially localized wave packet rather than a single plane wave, since the latter extends all over space and the question of tunneling times for the wave  is obscure. Due to the  spatial localization of the wave packet and the energy transfer from a spin, the wave packet is broadened in momentum space. Consequently, few modes with  energies higher than the barrier height can exist even when $U_0/E_0>1$ and the expectation value of the kinetic energy  $\langle E\rangle <U_0\pm \omega_0/2$ for the transmitted wave packet. The constructive interference between these modes and the modes with  energies lower than the barrier height form the wave packet   propagating through the barrier from left to right. When the wave packet arrives at the right end of the barrier and is transmitted out of the barrier, some modes get reflected at the boundary of the barrier. As the energy of the barrier increases, the higher energy modes are selectively transmitted at the boundary. Therefore, the parts of the wave packet which propagated non-adiabatically are preferentially transmitted, and  it appears that the probability of a spin-flip increases with the height of the barrier.

\begin{figure}
{%
\includegraphics[clip,width=\columnwidth]{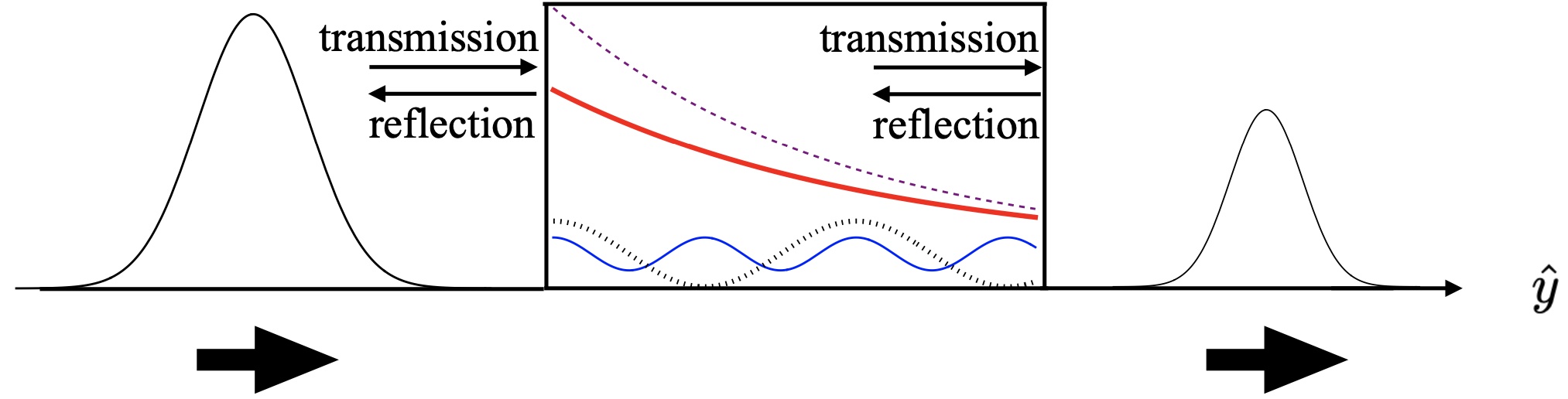} 
}
\caption{Wave packet propagating through the  barrier and examples of modes inside the barrier. The energies of the modes are lower than the barrier height for the dashed purple line and the thick red line, and they are higher than the barrier height for the dotted black line and the solid blue line. The dynamics of the wave packet inside the barrier is generally given by the sum of these modes. }
\label{fig6}
\end{figure}

\section{Conclusion}

To conclude, we numerically investigated the use of quantum clock for measuring time-of-flight and tunneling times. In particular, our study focused on the influence of the measurement-induced backaction and the filtering effect on  outcomes. We performed numerical simulations of measurements using the Larmor clock and the adiabatic theorem, respectively. It was observed that the Larmor tunneling time is shorter and the nonadiabatic transition probability of spin is larger for higher barriers. These results are consistent with each other and  with recent experimental result in \cite{exp,exp3}. Concerns about measuring time-of-flight and tunneling times with a quantum clock have  been the backaction due to the unavoidable energy transfer from a spin. Its strength is approximately equal to the inverse of the time resolution of the clock  \cite{time3}. Its effects  are mainly recorded in $\tau_z$ for the Larmor clock. In the case of the method using the adiabatic theorem, they give rise to the spin dependence of the nonadiabatic transition probability $\mathcal{P}$ and the transmission probability. We showed that $\mathcal{P}$ is always higher for the spin-up state than for the spin-down state due to the effects. Interestingly, it was observed that $\mathcal{P}$ increases with the height of the barrier in the tunneling regime for both initial states. This suggests that the shorter Larmor tunneling time and the larger $\mathcal{P}$ for higher barriers can be caused by the filtering effect. In the rectangular barrier, the filtering effect occurs at both the left and right edges. The high momentum modes are preferentially transmitted while low momentum modes are largely reflected as the wave packet enters and exits the barrier.  One of the ambiguities in the tunneling time problem is that the study of  time-dependent tunneling dynamics    generally requires the preparation of the spatially localized wave packet. This localization, together with the energy transfer from a spin, broadens the wave packet in momentum space. Therefore, rather than a single plane wave, it becomes important to investigate the time-dependent behavior of the constructive interference of modes within the wave packet. In particular, even if the wave packet consists mostly of modes with energies lower than the barrier height, there may exist few modes with energies higher than the barrier height for the reasons  above. The dynamics of the wave packet inside the barrier composed of these two types of modes is complex (Fig. 6). It can be expected that the wave packet traverses the barrier from left to right. Then the faster propagated parts  can be preferentially transmitted when exiting the barrier from the right end, resulting in the short Larmor tunneling time or large $\mathcal{P}$.  Note that the filtering effect at the right end of the barrier becomes pronounced when the barrier height approaches the energy of the particle and exceeds it to enter the tunneling regime. Since the transmitted wave packet can still consist mostly of modes with energies less than the barrier height, the expectation value of its energy can be less than the barrier height. The question remains how small the proportion of the modes with energies higher than the barrier height should be so that the dynamics of the wave packet can still be called quantum tunneling. Realistically, however, modes with energies lower and higher than the barrier height often coexist inside the barrier. Therefore  the study of the dynamics given by their sum may provide insight into tunneling time problems. In this paper, we numerically investigated two methods for measuring tunneling times using a quantum clock. Each method of measuring tunneling times has its inherent limitations. However, our study suggests that a comparison of outcomes from each method may clarify the origins of the behaviors observed and provide a deeper understanding of tunneling dynamics and measurements of it.

\begin{acknowledgements}
We thank W. H. Zurek, N. Sinitsyn, M. O. Scully, M. Arndt and C. H. Marrows for helpful discussions. F.S. acknowledges support from the Los Alamos National Laboratory LDRD program under project number 20230049DR and the Center for Nonlinear Studies. F.S. also thanks  the European Union’s Horizon 2020 research and innovation programme under the Marie Skłodowska-Curie Grant No. 754411 for support. W.G.U. thanks the Natural Science and Engineering Research Council of Canada, the
Hagler Institute of Texas A\&M Univ., the Helmholz Inst HZDR, Germany for
support while this work was being done.

\end{acknowledgements}

\end{document}